\documentclass[10pt,conference,a4paper]{IEEEtran}
\usepackage[latin1]{inputenc}
\usepackage[cmex10]{amsmath}
\usepackage{amsfonts}
\usepackage{amssymb}
\usepackage{graphicx}
\usepackage{url} %Embeds urls
\usepackage{hyperref} %Embeds urls
\usepackage{bm} % Easy way to make bold maths
\usepackage{enumerate} % Tools for lists
\usepackage{lhelp} % Tools for lists
\usepackage{listings}
\usepackage{color}

\title{Using Machine Learning to Decide When to Precondition Cylindrical Algebraic Decomposition With Groebner Bases}

\author{
	\IEEEauthorblockN{
	Zongyan Huang\IEEEauthorrefmark{1}, 
	Matthew England\IEEEauthorrefmark{2}, 
	James H. Davenport\IEEEauthorrefmark{3}, 
	and  
	Lawrence C. Paulson\IEEEauthorrefmark{1}
	}
	\IEEEauthorblockA{
	\IEEEauthorrefmark{1}University of Cambridge Computer Laboratory, Cambridge CB3 0FD, U.K.
	\\ Email:  \texttt{rubyhuang87@gmail.com}; \texttt{lp15@cam.ac.uk}}
	\IEEEauthorblockA{
	\IEEEauthorrefmark{2}Coventry University Faculty of Engineering, 
	Environment and Computing, Coventry, CV1 2JH, U.K.
	\\ Email: \texttt{Matthew.England@coventry.ac.uk}}
	\IEEEauthorblockA{
	\IEEEauthorrefmark{3}University of Bath Dept. of Computer Science, Bath, BA2 7AY, U.K.
	\\ Email: \texttt{J.H.Davenport@bath.ac.uk}}
}

\begin{document}

\maketitle

\begin{abstract} 
Cylindrical Algebraic Decomposition (CAD) is a key tool in computational algebraic geometry, particularly for quantifier elimination over real-closed fields.  However, it can be expensive, with worst case complexity doubly exponential in the size of the input.  Hence it is important to formulate the problem in the best manner for the CAD algorithm.  
One possibility is to precondition the input polynomials using Groebner Basis (GB) theory.  Previous experiments have shown that while this can often be very beneficial to the CAD algorithm, for some problems it can significantly worsen the CAD performance.  

In the present paper we investigate whether machine learning, specifically a support vector machine (SVM), may be used to identify those CAD problems which benefit from GB preconditioning.    We run experiments with over 1000 problems (many times larger than previous studies) and find that the machine learned choice does better than the human-made heuristic.
\end{abstract}

\section{Introduction}
\label{sec:intro}

\subsection{Cylindrical Algebraic Decomposition}
\label{ssec:introCAD}

A \emph{Cylindrical Algebraic Decomposition} (CAD) is a \emph{decomposition} of ordered $\mathbb{R}^n$ space into cells.  These are arranged \emph{cylindrically}, meaning the projections of any pair with respect to the given ordering are either equal or disjoint.  In this definition \emph{algebraic} is actually short for semi-algebraic as each CAD cell can be described with a finite sequence of polynomial constraints.  A CAD is produced to be invariant for input: sign- (or order-) invariant for input polynomials or truth-invariant for input formulae.

CADs and the first algorithm to compute them were introduced by Collins in 1975 \cite{Collins1975}.   CAD usually has two stages: projection where an operator is applied recursively on the input to derive corresponding problems in lower dimensions; and lifting where CADs are built incrementally by dimension according to the polynomials identified in projection \cite{ACM84I}.  The original motivation was quantifier elimination (QE) in real closed fields, while other applications include: 
parametric optimisation \cite{FPM05}, 
epidemic modelling \cite{BENW06}, 
theorem proving \cite{Paulson2012}, 
reasoning with multi-valued functions \cite{DBEW12}
derivation of optimal numerical schemes \cite{EH14}, and much more.

CAD has worst case complexity doubly exponential in the number of variables \cite{DH88} applicable whatever the data structure \cite{BD07}.  For some applications there exist algorithms with better complexity \cite{BPR06}, but CAD implementations still remain the best general purpose approach for many.  This may be due to the numerous approaches used to improve the efficiency of CAD since Collins' original work including: improvements to the projection operator \cite{Hong1990, McCallum1998, Brown2001a, HDX14}: partial CAD (lift only when necessary for QE) \cite{CH91}; and symbolic-numeric lifting schemes \cite{Strzebonski2006, IYAY09}.  Some recent advances include making use of any Boolean structure in the input \cite{BDEMW13, BDEMW16, EBD15}; local projection approaches \cite{Brown2013, Strzebonski2014a}; and decompositions via complex space \cite{CMXY09, BCDEMW14}.  For a more detailed introduction to CAD see for example Bradford et al. \cite{BDEMW16}.

\subsection{Preconditioning with Groebner Bases}
\label{ssec:introGB}

A \emph{Groebner Basis} $G$ is a particular generating set of an ideal $I$ defined with respect to a monomial ordering.  One definition is that the ideal generated by the leading terms of $I$ is generated by the leading terms of $G$.  Groebner Bases (GB) allow properties of the ideal to be deduced such as dimension and number of zeros and so are one of the main practical tools for working with polynomial systems.  Their properties and an algorithm to derive a GB for any ideal was introduced by Buchberger in his PhD thesis of 1965 \cite{Buchberger2006}.

Like CAD, there has been much research to improve and optimise GB calculation, with the $F_5$ algorithm \cite{Faugere2002} perhaps the most used approach currently.  However, also like CAD the calculation of GB is necessarily doubly exponential in the worst case \cite{MM82} (when using a lexicographic monomial ordering).  Despite this, the computation of GB can often be done very quickly and would almost certainly be a superior tool to CAD for any problem involving only polynomial equalities.  From this arises the natural question: is the process of replacing a conjunction of polynomial equalities in a CAD problem by their GB a useful precondition for CAD?  

I.e. let $E = \{e_1, e_2, \dots\}$ be a set of polynomials; 
$G = \{g_1, g_2, \dots\}$ a GB for $E$; and $B$ any Boolean combination of constraints, $f_i \, \sigma_i \, 0$, where $\sigma_i \in \{ <, >, \leq, \geq, \neq, =\}$) and $F = \{f_1, f_2, \dots\}$ is another set of polynomials.  Then %the two formulae formulae
\begin{align*}
\Phi &= (e_1 = 0 \land e_2 = 0 \land \dots) \land B \mbox{ and } \\
\Psi &= (g_1 = 0 \land g_2 = 0 \land \dots) \land B
\end{align*}
are equivalent and a CAD truth-invariant for either could be used to solve problems involving  $\Phi$ (such as eliminating any quantifiers applied to $\Phi$).  So is it worth producing $G$? % before calculating the CAD?

The first attempt to answer this question was given by Buchberger and Hong in 1991 \cite{BH91} who used the implementation of GB \cite{BGK85} to precondition an implementation of CAD \cite{CH91} (both in \textsc{C} on top of the \textsc{SAC-2} system \cite{Collins1985}).  Of the ten test problems studied: 6 were improved by the GB preconditioning, with the speed-up varying from 2-fold to 1700-fold; 1 problem resulted in a 10-fold slow-down; 1 timed out when GB preconditioning was applied, while it would complete without it; and the other 2 were intractable both for CAD alone and the GB preconditioning step.  

The problem was recently revisited by Wilson et al. \cite{WBD12_GB}.  The authors recreated the experiments of Buchberger and Hong \cite{BH91} using \textsc{Qepcad-B} for the CAD and \textsc{Maple 16} for the GB.  As we may expect, there had been a big decrease in the computation timings, especially the GB: the two test problems previously intractable \cite{BH91} could now have the GB calculated quickly.  However, two of the CAD problems were still hindered by GB preconditioning.
The experiments were then extended to: a wider example set (an additional 12 problems); the alternative CAD implementation in \textsc{Maple-16} \cite{CMXY09}; and the case where we further precondition by reducing inequalities of the system (the set $F$ above) with respect to the GB.  The key conclusion remained that GB preconditioning would in general benefit CAD (sometimes significantly) but could on occasion  hinder it (to the point of making a tractable CAD problem intractable).
The authors defined a metric to assist with the decision of when to precondition, the \emph{Total Number of Indeterminates} (\texttt{TNoI}) of a set of polynomials $A$,
\begin{equation}
\label{eq:TNoI}
\texttt{TNoI}(A) = \sum_{a \in A} \texttt{NoI}(a)
\end{equation}
where $\texttt{NoI}(a)$ is the number of indeterminates in a polynomial $a$.  Then their heuristic was to build a CAD for the preconditioned polynomials only if the \texttt{TNoI} decreased following preconditioning.  For most of their test problems the heuristic made the correct choice, but there were examples to the contrary and little correlation between the change in \texttt{TNoI} and level of speed-up / slow-down.

\subsection{Contribution and plan}
\label{ssec:contribution}

In this paper we consider whether machine learning can be applied to the decision of whether preconditioning CAD input with GB is beneficial for a particular problem.  We work on the reasonable assumption that GB computation is cheap for the problems on which CAD is tractable (in fact as shown in \cite{ED16} the CAD will compute resultants which overestimate the GB).  Hence we use algebraic features of both the input problem and the GB itself to decide \emph{whether we want to use} the GB.  

In Section \ref{sec:data} we describe the dataset and computer algebra computations used for the experiment and in Section \ref{sec:features} we describe the set of features identified to train the machine learning algorithm: a Support Vector Machine (SVM).  Then in Section \ref{sec:ML} we describe the initial machine learning experiment and its results, before running feature selection experiments in Section \ref{sec:featureselection}.  Finally, we compare the machine learned decision with the human developed \texttt{TNoI}-based heuristic, draw our conclusions and discuss future work in Section \ref{sec:summary}.

This is the second paper of the present authors to consider the application of SVMs to CAD optimisation.  We previously studied the choice of variable ordering for CAD in \cite{HEWDPB14}.  In that paper 3 existing heuristics were evaluated against a machine learned choice of which to use.  The latter outperformed each individually and suggested a greater role for machine learning in such decisions, motivating the present study.  The only other application of machine learning to computer algebra that the authors are aware of is by Kobayashi et al. \cite{KIMA16} who applied a SVM to decide the order of sub-formulae solving for QE.

\section{Dataset and computer algebra}
\label{sec:data}

\subsection{Computer Algebra}
\label{ssec-CA}

All the computer algebra computations were conducted in \textsc{Maple-17}.  The CAD algorithm used was an implementation of \cite{CMXY09}.  This is part of the \texttt{RegularChains} Library\footnote{\texttt{http://www.regularchains.org}} \cite{CM14d}, \cite{CM16} whose CAD procedures differ from the traditional projection and lifting framework of Collins, instead first decomposing $\mathbb{C}^n$ cylindrically and then refining to a CAD of $\mathbb{R}^n$.  Previous experiments \cite{WBD12_GB} showed this implementation has the same issues of GB preconditioning as the traditional approach.  The default \textsc{Maple} GB implementation was used: a meta algorithm calling multiple GB implementations.  The GBs were computed with a purely lexicographical ordering of monomials based on the same variable ordering as the CAD. 
 
All computations were performed on a 2.4GHz Intel processor, however this is not relevant as we evaluated the CAD performance using cell counts instead of timings, i.e. by comparing the numbers of cells in the final outputted CADs produced with and without GB preconditioning.  Numerous previous studies have shown this to be closely correlated to timings and it has the advantage of being discrete, machine and (up the theory used) implementation independent. It also correlates with the cost of any post-processing of the CAD.

\subsection{Dataset}
\label{ssec:dataset}

A key difficulty in applying machine learning techniques to computer algebra is the lack of suitable datasets.  CAD problem sets such as \cite{WBD12_EX} do not have anywhere near a sufficient number of problems to perform the experiment.  In our previous study on choosing the variable ordering for CAD \cite{HEWDPB14} we used the \texttt{nlsat}-dataset \cite{nlsat2012}, which although developed for non-linear arithmetic SAT-solvers, contained many suitable problems. 

For the present experiment we need problems that are expressed with a conjunction of at least two equalities in order to build a non-trivial GB.  From the nlsat dataset 493 three-variable problems and 403 four-variable problems fit this criteria, which should have been a sufficient number.  GB preconditioning was applied to each problem and cell counts from computing the CAD with the original polynomials and their replacement with the GB were computed and compared.  For each one of these problems the GB preconditioning was beneficial or made no difference; surprising as the experiments on much smaller datasets \cite{BH91, WBD12_GB} had shown much greater volatility.  This points to an undetected uniformity within the current nlsat dataset.  It would need to be widened if it is to be used more extensively for computer algebra research.

Since existing datasets were not suitable for the present experiment, we had no choice but to  generate our own problems.  The generation process aimed for an unbiased data set which would be computationally feasible for computing multiple CADs, and have some comparable structure (number of terms and polynomials) to existing CAD problems. 

In total, 1200 problems were generated using the random polynomial generator \texttt{randpoly} in \textsc{Maple-17}.  Each problem has two sets of three polynomials; the first to represent conjoined equalities and the second for the other polynomial constraints (respectively $E$ and $F$ from the description in Section \ref{ssec:introGB}).  The number of variables was at most $3$, labelled $x,y,z$ and under ordering $x \prec y \prec z$; the number of terms per polynomial at most 2; the coefficients were restricted to integers in $[-20,20]$; and the total degree was varied between 2, 3 and 4 (with 400 problems generated for each).  

A time limit of 300 CPU seconds was set for each CAD computation (all GB computations completed quickly) from which 1062 problems finished to constitute the final dataset.  Of these, 75\% benefited from GB preconditioning.   So our randomly generated dataset matched the previously found results of \cite{BH91, WBD12_GB} with most problems benefiting but not all and so is suitable for the purpose of this experiment.

%The usefulness of Groebner basis preconditioning was determined by whether it reduced the cell count or not, compared with a direct CAD.

%%%%%%%%%%%%%%%%%%%%%%%%%%%%%%%%%%%%%%%%%%%%%%%%%%%%%%%%%%%%%%%%%%%%%%%% 
%I don't think this code is needed given description in paragraph above.
%%%%%%%%%%%%%%%%%%%%%%%%%%%%%%%%%%%%%%%%%%%%%%%%%%%%%%%%%%%%%%%%%%%%%%%% 
%\begin{figure}[ht]
%\caption{Sample \textsc{Maple} command for random polynomial generation. The value of \texttt{num} varies between 2 and 4.}
%\begin{lstlisting}
%$E$ := [randpoly([x,y,z], terms = 2,
%    degree = num, coeffs = rand(-20 .. 20)),
%    randpoly([x,y,z], terms = 2,
%    degree = num, coeffs = rand(-20 .. 20)),
%    randpoly([x,y,z], terms = 2,
%    degree = num, coeffs = rand(-20 .. 20))
%    ];
%$B$ := [randpoly([x,y,z], terms = 2,
%    degree = num,coeffs = rand(-20 .. 20)),
%    randpoly([x,y,z], terms = 2,
%    degree = num,coeffs = rand(-20 .. 20)),
%    randpoly([x,y,z], terms = 2,
%    degree = num,coeffs = rand(-20 .. 20))
%    ];
%\end{lstlisting}
%\label{fig:randpoly}
%\end{figure}
%%%%%%%%%%%%%%%%%%%%%%%%%%%%%%%%%%%%%%%%%%%%%%%%%%%%%%%%%%%%%%%%%%%%%%%% 

\section{Problem features}
\label{sec:features}

Table \ref{table:28feature} shows the 28 problem features we identified (guided by previous work).  Here $(x,y,z)$ are the three variable labels used in the problems and proportion means the percentage of the total.  The features were chosen as easily computable metrics that may affect the cell count of the CAD.  They fall into two sets: those generated from the polynomials in the original problem and those obtained after applying GB preconditioning.  The abbreviations \texttt{tds} and \texttt{stds} stand for \emph{maximum total degree} and \emph{sum of total degrees} respectively:
\begin{align*}
\texttt{tds}(F) &= \max_{f \in F} \; \texttt{tds}(f),
\qquad 
\texttt{stds}(F) = \sum_{f \in F} \texttt{tds}(f).
\end{align*}
We also make use of the metric \texttt{TNoI} (see equation (\ref{eq:TNoI})) \cite{WBD12_GB}.  Finally we included the base 2 logarithm of the ratio of differences between some of the key metrics.  All features could be calculated immediately within \textsc{Maple}.

We note that the \texttt{stds} measure differs from the \texttt{sotd} heuristic introduced in \cite{DSS04} and used for multiple CAD decisions \cite{BDEW13, EBCDMW14, EBDW14}.  This is because \texttt{stds} measures the input polynomials only, while \texttt{sotd} measures the full set of CAD projection polynomials, and so is much more expensive.

In addition to training a classifier using all the features in Table \ref{table:28feature}, we trained classifiers using two subsets: one containing the features labelled $1-12$ concerning the original set of polynomials; and one the features labelled $13-25$ concerning the polynomials after GB preconditioning.  The latter set has one extra feature, the number of polynomials, as this varies after GB calculation but was always 6 to start with.  We refer to the first subset as \emph{before features}, the second as \emph{after features} and the full set as \emph{all features}.

\begin{table}[ht]
  \caption{Initial feature set}
  \label{table:28feature}
  \setlength{\tabcolsep}{10pt}
  \def\arraystretch{1}%
  \centering
    \begin{tabular}{c l} 
      \hline
        & Description \\  \hline \hline
      1 & \texttt{TNoI} before GB. \\ 
      2 & \texttt{stds} before GB. \\ 
      3 & \texttt{tds} of polynomials before GB.\\
      4 & Max degree of $x$ in polynomials before GB. \\
      5 & Max degree of $y$ in polynomials before GB. \\ 
      6 & Max degree of $z$ in polynomials before GB. \\
      7 & Proportion of polynomials with $x$ before GB. \\
      8 & Proportion of polynomials with $y$ before GB. \\ 
      9 & Proportion of polynomials with $z$ before GB.\\ 
      10 & Proportion of monomials with $x$ before GB.\\
      11 & Proportion of monomials with $y$ before GB.\\ 
      12 & Proportion of monomials with $z$ before GB.\\
      13 & Number of polynomials after GB.\\
      14 & \texttt{TNoI} after GB.\\
      15 & \texttt{stds} after GB.\\
      16 & \texttt{tds} of polynomials after GB.\\
      17 & Max degree of $x$ in polynomials after GB.\\
      18 & Max degree of $y$ in polynomials after GB.\\
      19 & Max degree of $z$ in polynomials after GB.\\
      20 & Proportion of polynomials with $x$ after GB. \\
      21 & Proportion of polynomials with $y$ after GB. \\ 
      22 & Proportion of polynomials with $z$ after GB.\\ 
      23 & Proportion of monomials with $x$ after GB.\\
      24 & Proportion of monomials with $y$ after GB.\\ 
      25 & Proportion of monomials with $z$ after GB.\\
      26 & $\log_2$(\texttt{TNoI} before GB) - $\log_2$(\texttt{TNoI} after GB)\\
      27 & $\log_2$(\texttt{stds} before GB) - $\log_2$(\texttt{stds} after GB)\\
      28 & $\log_2$(\texttt{tds} before  GB) - $\log_2$(\texttt{tds} after  GB) \\ \hline
    \end{tabular}
\end{table}

\noindent \textbf{Example:}  Consider sets of polynomials
\begin{align*}
E &:= \{
-12yz-3z,  \quad
17x^2-6,  \quad
-2yz+5x
\}
\\
F &:= \{
-2yz-9y,  \quad
-15x^2-19y,  \quad
6xz+3
\}.
\end{align*}
The GB computed for $E$ is
\[
G := \{
17x^2-6,  \quad
4y+1,  \quad
z+10x
\}
\]
and the \emph{all features} vector becomes
\begin{align*}
&\big[ 
\textstyle 12, 12, 2, 2, 1, 1, \frac{2}{3}, \frac{2}{3}, \frac{2}{3}, \frac{1}{3}, \frac{5}{12}, \frac{5}{12},\\
& \qquad \textstyle 6, 10, 10, 2, 2, 1, 1, \frac{2}{3}, \frac{1}{2}, \frac{1}{2}, \frac{1}{3}, \frac{1}{3}, \frac{1}{4},\\
& \qquad 0.263, 2.263, 0 
\big],
\end{align*} 
with the \emph{before features} and \emph{after features} vectors formed from the first and second line respectively.

The feature generation process was applied to create the three training sets separately (although the feature labels used were all as in Table \ref{table:28feature}). 
Each problem was labelled $+1$ if Groebner basis preconditioning is beneficial for CAD construction, or $-1$ otherwise. 
After feature generation the training data was standardised so each feature had zero mean and unit variance across the training set.  The same standardisation was then applied to features in the test set.

\section{Machine learned choices}
\label{sec:ML}

\subsection{Introduction}
\label{ssec:MLIntro}

Machine learning deals with the design of programs that learn rules from data.  This is an attractive alternative to manually constructing them when the underlying functional relationship is complex, as appears to be the case here. 
%Machine learning techniques have been widely used in many fields, such as web searching\cite{boyan1996machine}, text categorization\cite{sebastiani2002machine}, robotics\cite{stone2000multiagent}, expert systems\cite{forsyth1986machine} and many others.
%Various machine learning techniques have been developed. McCulloch and Pitts\cite{mcculloch1943logical} created the first computational model for \emph{neural networks} called \emph{threshold logic}. Following that, Rosenblatt~\cite{rosenblatt1958perceptron} proposed the \emph{perceptron} as an iterative algorithm for supervised classification of an input into one of several possible non-binary outputs. A later development was the \emph{decision tree} \cite{alpaydin2004introduction}, which is a simple representation for classifying examples. The main idea here is to apply serial classifications which refine the output state. At the same time as the \emph{decision tree} was being developed, the \emph{multi-layer perceptron} \cite{hornik1989multilayer} was explored. It is a modification of the standard linear perceptron and can distinguish data that are non-linearly separable. 

In the last decade, the use of machine learning has spread rapidly following the invention of the \emph{Support Vector Machine} (SVM) (see for example \cite{STV04}). 
%This was a development of the perceptron approach and 
This gives a powerful and robust method for both: \emph{Classification}, the assignment of input examples into a given set of classes; and \emph{Regression}, a supervised pattern analysis in which the output is real-valued.  
%The SVM technology can deal efficiently with high-dimensional data, and is flexible in modelling diverse sources of data. 
The standard SVM classifier takes a set of input data and predicts one of two possible classes from the input. Given a set of examples, each marked as belonging to one of two classes, an SVM training algorithm builds a model that assigns new examples into one of the classes. The examples used to fit the model are called training examples. 
An important concept in the SVM theory is the use of a kernel function to map data into a high dimensional feature space and then separate samples in the transformed space \cite{SC04}. Kernel functions enable operations in feature space without ever computing the coordinates of the data in that space, rather they compute the inner products between all pairs of data vectors, which is generally cheaper.
%This operation is generally computationally cheaper than the explicit computation of the coordinates. 

For our experiment we used \textsc{SVM-Light}\footnote{\url{http://svmlight.joachims.org}} \cite{Joachims1999}; an implementation of SVMs in \textsc{C}. 
%The \textsc{SVM-Light} software consists of two programs: \textsc{SVM learn} and \textsc{SVM classify}. \textsc{SVM learn} fits the model parameters based on the training data and user inputs (such as the kernel function and the parameter values). \textsc{SVM classify} uses the generated model to classify new samples.
%It calculates a hyperplane of the $n$-dimensional transformed feature space, which is an affine subspace of dimension $n - 1$ dividing the space into two corresponding to the two distinct classes.  \textsc{SVM classify} outputs margin values which are a measure of how far the sample is from this separating hyperplane. Hence the margins are a measure of the confidence in a correct prediction. A large margin represents high confidence in a correct prediction. The accuracy of the generated model is largely dependent on the selection of the kernel functions and parameter values. 

\subsection{Cross-validation and grid-search}
\label{ssec:cvgs}

The 1062 problems were partitioned into 80\% training (849 problems) and 20\% test (213 problems), stratified to maintain relative proportions of positive and negative examples. 
The classification was done %in \textsc{SVM-Light} 
using the \emph{radial basis function} (RBF) kernel.  This was chosen after earlier experiments applying machine learning to an automated theorem prover found the RBF kernel to perform well with similar simple algebraic features \cite{BHP14}.  
The RBF function is defined as:
\begin{equation}
\label{eq:RBF}
K(x, x\prime ) = \exp \left( -\gamma ||x - x\prime ||^2 \right)
\end{equation}
where $x$ and $x\prime$ are feature vectors.  The process depends on kernel parameter $\gamma$ and another parameter $C$ which governs the trade-off between margin and training error, and finding the optimal values of these is not trivial.  \emph{Matthews' Correlation Coefficient} (MCC) \cite{Matthews1975, BBCAN00} is often used to evaluate choices.  This takes into account true and false positives and negatives (labelled tp, fp, tn and fn accordingly):
\begin{equation}
\label{eq:MCC}
{\rm MCC} = \frac{{\rm tp}*{\rm tn}-{\rm fp}*{\rm fn}}{\sqrt{({\rm tp}+{\rm fp})({\rm tp}+{\rm fn})({\rm tn}+{\rm fp})({\rm tn}+{\rm fn})}}.
\end{equation}
In the case where one of the terms in the denominator is zero the entire denominator is set to 1.  The MCC measure has the value $1$ if perfect prediction is attained, $0$ if the classifier is performing as a random classifier, and $-1$ if the classifier exactly disagrees with the data. 

A grid-search optimisation procedure along with a five-fold stratified cross validation was used, involving a search over a range of $(\gamma ,C)$ values to find the pair which would maximize equation (\ref{eq:MCC}).  We tested a commonly used range of values in our grid search process \cite{HCL03}: $\gamma$ varied between $\{2^{-15}, 2^{-14}, 2^{-13}, \dots, 2^{3}\}$; and $C$ varied between $\{2^{-5}, 2^{-4},$ $2^{-3}, \dots, 2^{15}\}$. 
Following the completion of the grid-search, the values giving optimal MCC results were selected.  This procedure was repeated for the three feature sets.

\subsection{Results for the three feature sets}
\label{ssec:results}

%I compared the machine learning outcomes, between results obtained with \emph{all features} used, with just the \emph{before features}, and with just the \emph{after features}. 
The classification accuracy was used to measure the efficacy of the machine learning selection process under the 3 feature sets.  The test set of 213 problems contained 159 positive samples and 54 negative samples (i.e. 75\% of the test problems benefited from GB preconditioning). 
%Thus we may consider this a baseline for our machine learned decisions.
%In 75\% of the cases Groebner basis preconditioning was beneficial for CAD construction for the given problem; 
%this was used as a baseline for measuring the efficacy of the classifiers. 
The results of the machine learned choices are summarised in Table~\ref{tab:numof}.
First we note that when making a choice based on the \emph{before features} training set 75\% of the problems were predicted accurately. I.e. making a decision based on these features results in no more correct decisions than blindly deciding to GB precondition each and every time.  However, the other two feature sets resulted in superior decisions.  
Although only a small improvement on preconditioning blindly, we recall that the wrong choice can give large changes to the size of the CAD or even change the tractability of the problem \cite{WBD12_GB}. 

The results indicate that the features of the GB itself are required to decide whether to use the preconditioning.  
However, we cannot conclude this directly: earlier research shows that a variable completely useless by itself can provide a significant performance improvement when taken in conjunction with others \cite{GE03}.  To be confident about which features were significant and which were superfluous, further feature selection experiments are required and we will see that the optimal feature subset must contain features from both before and after the GB computation.

\begin{table}[ht]
  \caption{Accuracy of predictions}
  \label{tab:numof}
\centering
   \def\arraystretch{1.2}%
    \begin{tabular}{lcc}
      \hline
      Feature Set 		 	& Number 	& \% of test set\\  \hline \hline
%      Always applying GB 	& 159 		& 75\%  \\
      All features 		 	& 162 		& 76\% \\
      Before features 	 	& 159 		& 75\% \\
      After features 	 	& 167 		& 78\% \\ \hline
    \end{tabular}
\end{table}

\section{Feature selection}
\label{sec:featureselection}

%Given the fact that \emph{after features} is a subset of \emph{all features}, there is an indication that not all the features contribute to the machine learning process, only a small number of features are needed for learning to be effective. Moreover, 
There is a strong indication that not all features contribute to the machine learning process.  Moreover, a reduced feature set can be beneficial for better understanding the underlying connections. Consequently, we applied some \emph{feature selection} methods.  Both filter and wrapper methods were applied as discussed in the following subsections. 

The feature selection experiments were conducted with \textsc{Weka} (Waikato Environment for Knowledge Analysis) \cite{HFHPRW09}, a Java machine learning library which supports tasks such as data preprocessing, clustering, classification, regression and feature selection. Each data point is also represented as a fixed number of features. The inputs are samples of $29$ features, where the first $28$ are the real-valued features from Table \ref{table:28feature}, and the final one is a \emph{nominal} feature denoting its class. 

\subsection{The filter method}
\label{ssec:filter}

A correlation based feature selection method, was applied as described in \cite{HH03}. Unlike other filter methods \cite{Hall2000}, these measure the rank of feature subsets instead of individual features.  A feature subset which contains features highly correlated with the class but uncorrelated with each other is preferred.  The metric below is used to measure the quality of a feature subset, and takes into account feature-class correlation as well as feature-feature correlation. 
\begin{equation}
\label{eq:Gs}
G_s = \frac{k\overline{r_{ci}}}{\sqrt{k+k(k-1)\overline{r_{ii^{'}}}}}
\end{equation}
Here, $k$ is the number of features in the subset, $\overline{r_{ci}}$ denotes the average feature-class correlation of feature $i$, and $\overline{r_{ii^{'}}}$ the average feature-feature correlation between feature $i$ and $i'$. 
The numerator of equation (\ref{eq:Gs}) indicates how much relevance there is between the class and a set of features, while the denominator measures the redundancy among the features. The higher $G_s$, the better the feature subset. 

%In order to apply this heuristic to estimate the merit of a feature subset, it is necessary to compute the feature-class correlations and feature-feature correlations. 
To apply this heuristic we must calculate the correlations.
With the exception of the class attribute all 28 features are continuous, so in order to have a common measure for computing the correlations we first discretize using the method of Fayyad and Irani \cite{FI93}. After that, a correlation measure based on the information-theoretical concept of \emph{entropy} is used, which is a measure of the uncertainty of a random variable.  We define the \emph{entropy of a variable} $X$  \cite{Shannon2001} as 
\begin{equation}
H(X)=-\sum_{i} p(x_i)\log_2\big(p(x_i)\big).
\end{equation}
The entropy of $X$ after observing values of another variable $Y$ is then defined as
\begin{equation}
\textstyle H(X|Y) = -\sum_{j} p(y_j) \sum_{i} p(x_i|y_j) \log_2\big(p(x_i|y_j)\big),
\end{equation}
where $p(x_i)$ is the prior probabilities for all values of $X$, and $p(x_i|y_i)$ is the posterior probabilities of $X$ given the values of $Y$.
The \emph{information gain (IG)} \cite{Quinlan1986} measures the amount by which the entropy of $X$ decreases by additional information about $X$ provided by $Y$, and it is given by
\begin{equation}
IG(X,Y) = H(X) - H(X|Y). 
\end{equation}
The \emph{symmetrical uncertainty (SU)} (a modified information gain measure) is then used to measure the correlation between two discrete variables (X and Y) \cite{PTVF92}: 
\begin{equation}
SU(X,Y)=2.0 \times \bigg(\frac{H(X) - H(X|Y)}{H(X)+H(Y)}\bigg).
\end{equation}

Treating each feature as well as the class as random variables, we can apply this as our correlation measure. More specifically, we simply use $SU(c,i)$ to measure the correlation between a feature $i$ and a class $c$, and $SU(i,i^{'})$ to measure the correlation between features $i$ and $i'$. These values are then substituted as $\overline{r_{ci}}$ and $\overline{r_{ii^{'}}}$ in equation (\ref{eq:Gs}). 

Recall that our aim here is to find the optimal subset of features which maximises the metric given in equation (\ref{eq:Gs}). 
The size of our feature set is 28 meaning there are
%\[
$
2^{28}-1 \simeq 2.7 \times 10^8
$
%\]
possible subsets, too many for exhaustive search. 
Instead a greedy stepwise forward selection search strategy was used for searching the space of feature subsets, which works by adding the current best feature at each round. The search begins with the empty set, and in each step the metric, as defined in equation (\ref{eq:Gs}), is computed for every single feature addition, and the feature with the best score improvement is added.  If at some step none of the remaining features provide an improvement, the algorithm stops, and the current feature set is returned.  The best feature subset found with this method (which may not be the absolute optimal subset of features) is shown in Table \ref{tab:filter}, ordered by importance. 

\begin{table}[ht]
  \caption{Feature selection by the filter method}
  \label{tab:filter}
\centering
\def\arraystretch{1.2}%
    \begin{tabular}{c l}
      \hline
       		& Description \\  \hline \hline
      14 	& \texttt{TNoI} after GB.\\
      13 	& Number of polynomials after GB.\\
      2 	& \texttt{stds} before GB.  \\
      26 	& lg(\texttt{TNoI} before GB) - lg(\texttt{TNoI} after GB)\\ 
      21 	& Proportion of polynomials with $y$ after GB.\\
      15 	& \texttt{stds} after GB.\\
      23 	& Proportion of monomials with $x$ after GB.\\
      19 	& Max degree of $z$ in polynomials after GB.\\
      25 	& Proportion of monomials with $z$ after GB.\\
      27 	& $\log_2$(\texttt{stds} before GB) - $\log_2$(\texttt{stds} after GB)\\ \hline
    \end{tabular}
\end{table}

\subsection{The wrapper method} 
\label{ssec:wrapper}

The wrapper feature selection method evaluates attributes using accuracy estimates provided by the target learning algorithm. Evaluation of each feature set was conducted with a learning scheme (a SVM with RBF kernel function). The SVM algorithm is run on the dataset, with the same data partitions as described in Section \ref{ssec:cvgs}. Similarly, a five-fold cross validation was carried out. The feature subset with the highest average accuracy was chosen as the final set on which to run the SVM algorithm. 

In each training / validation fold, starting with an empty set of features: each feature was added; a model was fitted to the training data set; the classifier was then tested on the validation set. This was done on all the features, resulting in a score for each where the score reflects the accuracy of the classifier. The final score for each feature was its average over the five folds. Having obtained a score for all features in the manner above, the feature with the highest score was then added in the feature set. Then, the same greedy procedure as described for the filter method in Section \ref{ssec:filter} was applied to obtain the best feature subset.   

Due to the large number of cases, the parameters $(C, \gamma)$ were selected from an optimised sub range instead of the full grid search used in Section \ref{ssec:cvgs}.  The reduced range suffices to demonstrate the performance of a reduced feature set. In those previous experiments %with \emph{all features}, \emph{before features} and \emph{after features} 
we found that $C$ taken from $\{2^{5}, 2^{6}, 2^{7}, 2^{8}, 2^{9}, 2^{10}\}$ and $\gamma$ taken from $\{2^{-5}$, $2^{-6}$, $2^{-7}, 2^{-8}, 2^{-9}, 2^{-10}\}$ provided good classifier performance.
% (the best parameter settings from \emph{all features}, \emph{before features} and \emph{after features} all fell into this range). 

The 36 pairs of $(C, \gamma)$ values were tested and an optimal feature subset with the highest accuracy was found for each. Then the one with the highest accuracy was selected as the final set, which is shown in Table \ref{tab:wrapper} ordered by importance.  We see that most of the features selected (9, 12 and 22) related to variable $z$. 
Recall that the projection order used in the CAD was always $x \prec y \prec z$, i.e. the variable $z$ is projected first.  Hence it makes sense that this variable would have the greatest effect and thus be identified in the feature selection.

\begin{table}[t]
  \caption{Feature selection by the wrapper method}
  \label{tab:wrapper}
\centering
    \begin{tabular}{c l}
      \hline
      		& Description \\  \hline \hline
      14 	& \texttt{TNoI} after GB.\\
      9 	& Proportion of polynomials with $z$ before GB.\\ 
      22 	& Proportion of polynomials with $z$ after GB. \\
      4 	& Max degree of $x$ in polynomials before GB. \\
      12 	& Proportion of monomials with $z$ before GB.\\ \hline
    \end{tabular}
\end{table}

\begin{figure}[t]
\centering
  \caption{Performance of a sample run with different sizes of feature sets}
  \label{fig:performance}
\includegraphics[width=0.4\textwidth]{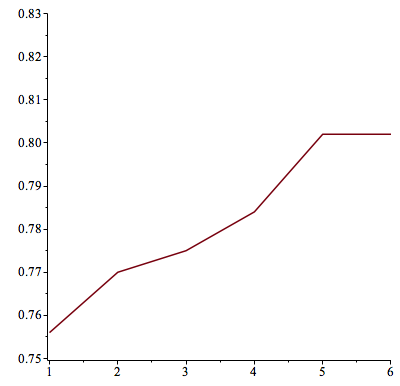}
\end{figure}

We examined the performance on further reduced feature sets, obtained by the feature ranking of the wrapper method. Figure \ref{fig:performance} shows the overall prediction accuracies. For instance, the predictor obtained from only using a single feature (the best ranked feature was TNoI after GB for both filter and wrapper methods) achieved an accuracy score of 0.756 in that run, with the performance steadily increasing with the size of the feature set until the fifth feature. Taking any sixth feature into the set did not improve the performance noticeably, and hence resulted in the cut-off chosen by the wrapper method.

\begin{figure}[t]
\begin{center}
  \caption{Feature 4 and 14 of misclassified data}
  \label{fig:error1}
\includegraphics[width=0.45\textwidth]{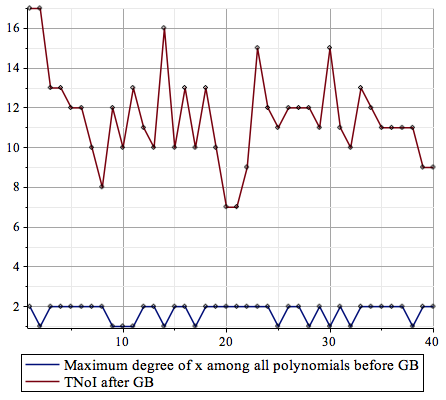}
\end{center}
\end{figure}

\begin{figure}[t]
\begin{center}
  \caption{Feature 9,12 and 22 of misclassified data}
  \label{fig:error2}
\includegraphics[width=0.48\textwidth]{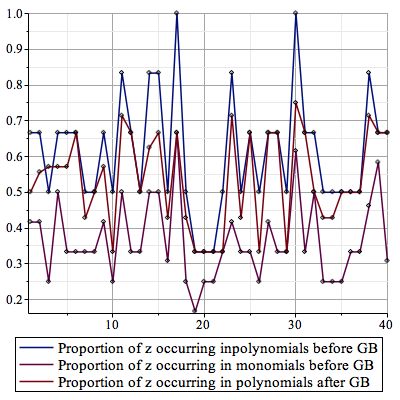}
\end{center}
\end{figure}

As the wrapper method identified only a few features an error analysis on the misclassified data points is feasible. Figure \ref{fig:error1} shows 40 misclassified points and their features 4 and 14, while Figure \ref{fig:error2} shows the remaining features. 
% 9, 12, 22 of the same samples. 
It is interesting that feature 4 of all misclassified samples is either 1 or 2, when for the whole data set roughly a third of samples had this feature value 3 or 4.  This indicates that the algorithm performs better on instances with a higher maximum degree of $x$ among all polynomials before GB preconditioning.

\subsection{Results with reduced feature sets}
\label{ssec:featureresults}

Having obtained the reduced feature sets, we ran the experiment again to evaluate the new choices.
The data set was again repartitioned into 80\% training and 20\% test set, stratified to maintain relative class proportions in both training and test partitions. Again, a five-fold cross validation and a finer grid-search optimisation procedure over the range of ($C$, $\gamma$) pairs was conducted as described previously.
%where $\gamma$ varies between $2^{-15}, 2^{-14}, 2^{-13}, \dots, 2^{3}$ and $C$ varies between $2^{-5}, 2^{-4}, 2^{-3}, \dots, 2^{15}$ (as described in Section xyz) to determine the parameter selection, and the selected features were used. 
The classifier with maximum averaged MCC %\cite{Matthews1975, BBCAN00} 
was selected and the resulting classifier was then evaluated. The testing data was also reduced to contain only the features selected.  The classification accuracy was used to measure the performance of the classifier. In order to better estimate the generalisation performance of classifiers with reduced feature sets, the data was permuted and partitioned into 80\% training and 20\% test again and the whole process was repeated 50 times. For each run, each training set was standardised to have zero mean and unit variance, with the same offset and scaling applied subsequently to the corresponding test partition. 

\begin{figure}[ht]
\begin{center}
  \caption{Boxplots of 50 runs of the 5-fold cross validation with both the suggested feature sets}
  \label{fig:featurebox}
\includegraphics[width=0.48\textwidth]{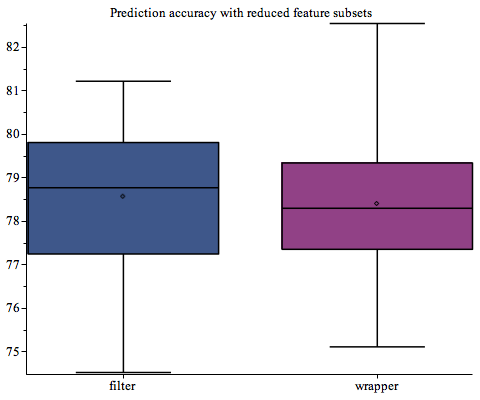}
\end{center}
\end{figure}

Figure \ref{fig:featurebox} shows boxplots of the accuracies generated by 50 runs of the five-fold cross validation. Both reduced feature sets generated similar results and show a large improvement on the base case where Groebner basis preconditioning is always used before CAD construction. 
The average overall prediction accuracy of the filter subset and the wrapper subset is 79\% and 78\% respectively (we note that Figure \ref{fig:performance} shows a higher rate but that was just for one sample run).  All 50 runs of the wrapper subset performed above the base line, while the top three quartiles of the results of both sets achieve higher than 77\% percentage accuracy.   

\section{Conclusion and future work}
\label{sec:summary}

\subsection{Comparison with human developed heuristic}

We may compare the machine learned choice with the human developed \texttt{TNoI} heuristic \cite{WBD12_GB}, whose performance on the 213 test problems is shown in Table \ref{tab:TNoI}.  It correctly predicted whether GB preconditioning was beneficial for 118 examples, only 55\%.  So for this dataset it would have been better on average to precondition blindly than to make a decision using \texttt{TNoI} alone.  
The \texttt{TNoI} heuristic performed better in the experiments by Wilson et al. \cite{WBD12_GB}.  Those experiments involved only 22 problems (compared to 213 in the test set here) but they were human constructed to have certain geometric properties while the ones here were random.  

We also note that the \texttt{TNoI} heuristic actually performed quite differently for positive and negative examples of our dataset, as shown by the separated data in Table \ref{tab:TNoI}.  It was able to identify most of the cases where GB-preconditioning is detrimental but failed to identify many of the cases where it was beneficial.  The \texttt{TNoI} after GB was identified as important by both feature methods, but it seems to need to be used in conjunction with other features to be effective here.

\subsection{Summary}

We investigated the application of machine learning to the problem of predicting when GB preconditioning is beneficial to CAD.  We had to create a new dataset of random polynomials for the experiment.  We acknowledge that it would be preferable to run such a test on an established dataset but this was not available.  Of course, such a random datasets could be enlarged to increase variety almost indefinitely, but we needed to keep the experiment within computationally feasible boundaries.  We emphasise the interesting initial finding in Section \ref{ssec:dataset} that supposedly varied established sets can have hidden uniformity; and highlight that our generated dataset matched previously reported results \cite{BH91, WBD12_GB} for the topic of study with most, but not all, benefiting from GB preconditioning.

A machine learned choice on whether to precondition was found to yield better results than either always preconditioning blindly, or using the previously human developed \texttt{TNoI} heuristic \cite{WBD12_GB}.
Two feature selection experiments showed that a small feature subset could be used. The two subsets identified were different but both needed features from before and after the GB preconditioning.  For one, having fewer features actually improved the learning efficiency to 79\%.  Although a modest improvement on applying GB preconditioning blindly, we recall that the wrong choice can give large changes to the size of the CAD or even change the tractability of the problem. 

\begin{table}[t]
  \caption{The performance of the \texttt{TNoI}-based heuristic \cite{WBD12_GB}}
  \label{tab:TNoI}
\centering
    \begin{tabular}{lccc}
      \hline
      					& Total & Correct Prediction  \\
      \hline \hline
      GB beneficial 	& 158   & 77  & (48\%) \\
      GB not beneficial & 54    & 41  & (76\%) \\
      Total             & 213   & 118 & (55\%) \\
      \hline
    \end{tabular}
\vskip-0.18in
\end{table}

\subsection{Future Work}

\noindent There are many possible extensions to this project:
\begin{itemize}
\item To see how the learned choice performs on a non-random, dataset.  There is a large set derived from university mathematics entrance exams \cite{KIMA16}, which is not yet publicly available but may be in the future.
\item There are further CAD optimisations for multiple equalities under development \cite{EBD15, ED16, DE16} which may affect the role of GB preconditioning from CAD.
\item In the present paper the variable ordering for CAD and the monomial ordering for GB were fixed.  In reality, such decisions may also need to be made in tandem and it is an open problem as to how best to do this.  The variable ordering can affect the choice of whether to use GB preconditioning and \emph{vice versa}.
%, preconditioning could change the best choice of variable ordering. 
%It is possible that machine learning could be used to make such tandem choices.
\end{itemize}
Finally, we note there are other algorithm optimisation decisions for CAD, and indeed elsewhere in computer algebra.

\subsection*{Acknowledgements}

%Thanks to the referees for their helpful comments.
Thanks to David Wilson and James Bridge, our collaborators on \cite{HEWDPB14}, for useful conversations on the topic of machine learning to optimise computer algebra.  
This work was supported by EPSRC grant EP/J003247/1 and EU H2020-FETOPEN-2016-2017-CSA project $\mathcal{SC}^2$ (712689).

\end{document}